\documentclass[prl,aps,twocolumn,superscriptaddress,showpacs]{revtex4}
\usepackage{amsmath,amssymb,graphicx}
\usepackage{pxfonts}
\usepackage{graphicx}
\usepackage{color}
\usepackage{float}
\begin{document}

\title{Theory of condensation of indirect excitons in a trap}

\author{S.V. Lobanov}
\affiliation{Skolkovo Institute of Science and Technology, Moscow 143026, Russia}
\affiliation{School of Physics and Astronomy, Cardiff University, Cardiff CF24 3AA, United Kingdom}

\author{N.A. Gippius}
\affiliation{Skolkovo Institute of Science and Technology, Moscow 143026, Russia}

\author{L.V. Butov}
\affiliation{Department of Physics, University of California at San Diego, La Jolla, California 92093-0319, USA}

\begin{abstract}
We present theoretical studies of condensation of indirect excitons in a trap. Our model quantifies the effect of screening of the trap potential by indirect excitons on exciton condensation. The theoretical studies are applied to a system of indirect excitons in a GaAs/AlGaAs coupled quantum well structure in a diamond-shaped electrostatic trap where exciton condensation was studied in earlier experiments. The estimated condensation temperature of the indirect excitons in the trap reaches hundreds of milliKelvin.
\end{abstract}

\pacs{73.63.Hs, 78.67.De}

\date{\today}

\maketitle

\section{I. Introduction}

Potential traps made possible the realization of Bose-Einstein condensation of atoms \cite{Cornell02,Ketterle02}. Traps also became an effective tool for studying cold excitons -- cold bosons in condensed matter materials. The realization of a cold and dense exciton gas in a trap requires long exciton lifetimes allowing excitons to travel to the trap center and cool to low temperatures before recombination. Furthermore, the realization of a cold and dense exciton gas requires an excitonic state to be the ground state \cite{Keldysh86}. These requirements are fulfilled in a system of indirect excitons (IXs). An IX in a
coupled quantum well structure (CQW) is composed of an electron and a hole in spatially separated layers (Fig.~1a). The spatial separation allows one to control the overlap of electron and hole wave functions and engineer structures with lifetimes of IXs orders of magnitude longer than lifetimes of regular excitons \cite{Lozovik76, Fukuzawa90}.

\begin{figure}
\includegraphics{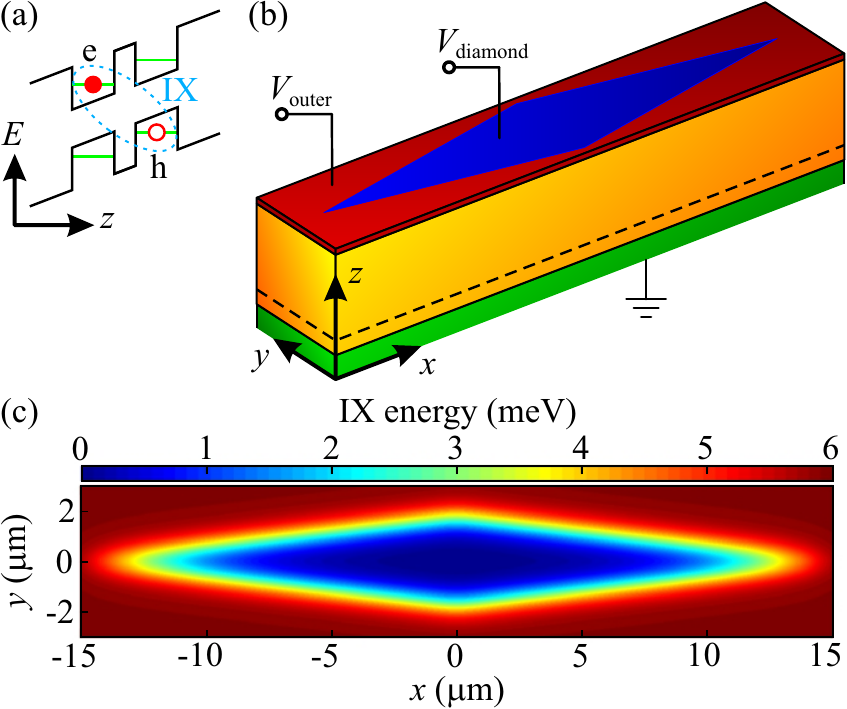} %[width=8.5cm]
\caption{(a) Energy band diagram of the CQW; e, electron; h, hole. The cyan dashed ellipse indicates an indirect exciton (IX). (b)~Schematic of device. CQW (black dashed lines) is positioned within an undoped 1~$\mu$m thick Al$_{0.33}$Ga$_{0.67}$As layer (yellow) between a conducting $n^+$-GaAs layer serving as a homogeneous bottom electrode (green) and a $3.5 \times 30$~$\mu$m diamond-shaped top electrode (blue) surrounded by an 'outer plane' electrode (red). (c)~Simulation of IX energy profile in the bare trap for $V_{\rm diamond} = -2.5$~V and $V_{\rm outer} = -2$~V.}
\end{figure}

In a set of experimental studies, IXs were created in a GaAs/AlGaAs CQW. Long lifetimes of the IXs allow them to cool to low temperatures within about 0.1~K of the lattice temperature a few nanoseconds after the generation \cite{Butov01} or a few micrometers away from the excitation spot \cite{Hammack06PRL}. In turn, the lattice temperature can be lowered to about 50~mK in an optical dilution refrigerator. This allows the implementation of a cold exciton gas with temperatures well below the temperature of quantum degeneracy $T_\mathrm{d} = 2 \pi \hbar^2 n/(k_\mathrm{B}m)$ (for a typical GaAs CQW with the exciton mass $m = 0.22 m_0$, $T_\mathrm{d} \sim 3$~K for the exciton density per spin state $n = 10^{10}$~cm$^{-2}$).

A trapping potential for IXs can be created by voltage. IXs have a built-in dipole moment $ed$, where $d$ is the separation between the electron and hole layers. An electric field $F_z$ perpendicular to the QW plane results in the IX energy shift $E = - edF_z$ \cite{Miller85}. This gives the opportunity to create in-plane potential landscapes for IXs $E(x,y)= - edF_z(x,y)$. Advantages of electrostatically created potential landscapes include the opportunity to realize a desired in-plane potential profile and control it by voltage in situ, i.e. on a time scale shorter than the IX lifetime. IXs were studied in a variety of electrostatic traps \cite{Huber98, Hammack06JAP, Gorbunov06, Chen06, Rontani09, High09NL, High09PRL, Schinner11, Kowalik-Seidl12, High12, Gorbunov12, Gorbunov13, Schinner13PRL, Schinner13PRB, Shilo13, Kuznetsova15, Cohen16} as well as in periodic trap potentials including static \cite{Zimmermann97, Zimmermann98, Remeika09, Remeika12, Remeika15} and moving \cite{Winbow11, Hasling15} electrostatic lattices.

Other potential landscapes for IXs that can be controlled in situ include moving lattices created by surface acoustic waves \cite{Rudolph07, Lasic10, Lasic14, Violante14} and laser-induced traps \cite{Hammack06PRL, Gorbunov11, Alloing13}. Excitons were also studied in a variety of traps whose profile cannot be changed in situ. These traps include strain-induced traps \cite{Trauernicht83, Kash88, Negoita99, Naka05, Yoshioka11}, traps created by laser-induced interdiffusion \cite{Brunner92}, and magnetic traps~\cite{Christianen98}.

The bosonic nature of excitons allows for their Bose-Einstein condensation at low temperatures \cite{Keldysh68}. Spontaneous coherence and condensation of IXs was measured in a diamond-shaped electrostatic trap \cite{High12}. In this work, we present theoretical studies of IX condensation in the trap.

\section{II. Theoretical model}

The trap is created using a diamond-shaped top electrode (Fig.~1b). Because a thinner electrode produces a smaller $F_z$ due to field divergence near the electrode edges, a voltage between the diamond-shaped electrode and homogeneous bottom electrode creates a confining potential with the IX energy gradually reducing toward the trap center \cite{High09PRL}. The considered device includes a $3.5 \times 30$~$\mu$m diamond electrode at $V_{\rm diamond} = -2.5$~V surrounded by an 'outer plane' electrode at $V_{\rm outer} = -2$~V. Two 8 nm GaAs QWs separated by a 4 nm Al$_{0.33}$Ga$_{0.67}$As barrier are positioned 100 nm above the $n^+$-GaAs layer within an undoped 1~$\mu$m thick Al$_{0.33}$Ga$_{0.67}$As layer. Positioning the CQW closer to the homogeneous electrode suppresses the in-plane electric field \cite{Hammack06JAP}, which otherwise can lead to exciton dissociation. This configuration corresponds to the experimental system studied in Ref.~\cite{High12}. The IX energy profile in the bare trap calculated numerically from the Poisson equation is presented in Fig.~1c. A confining potential of the diamond trap allows collecting IXs from a large area, facilitating the creation of a dense and cold IX gas in the trap and, in turn, exciton condensation in the trap. Furthermore, as shown below a parabolic-like confining potential of the diamond trap increases the condensation temperature.

IXs in the trap are considered within the following approximation:

(i) The dimensions of the trap are much larger than the IX Bohr radius, which is $\sim 10$~nm in the structure, and the potential energy variation on the IX Bohr radius is negligible in comparison to the IX binding energy, which is $\sim 4$~meV~\cite{Szymanska03, Sivalertporn12}. Therefore, IXs are considered as point-like Bose-particles within the model. Exact two-body quantum dynamics of IXs in electrostatic potentials was studied in Refs.~\cite{Grasselli15, Grasselli16}.

(ii) Interactions play a key role in the physics of IX systems. In the first approximation, the exciton-exciton interaction potential is given by $v(r) = \frac{2 e^2}{\kappa} \left(\frac{1}{r} - \frac{1}{\sqrt{r^2 + d^2}}\right)$, where $d$ is the center-to-center separation of the two wells in the CQW and $\kappa$ is the dielectric constant of the semiconductor. At $r \gg d$ this potential has the form of the dipole-dipole repulsion, $v(r) \simeq e^2 d^2 / (\kappa r^3)$. Interaction between IXs lead to an increase of IX photoluminescence (PL) energy $E$ with density $n$, which has been known since early spectroscopic studies of IXs~\cite{Butov94, Butov99}. The Hartree approximation of the exciton interaction gives the PL energy shift
\begin{equation}
\Delta E_\mathrm{c} = n \int v({r}) d^2 r = \frac{4\pi e^2 d}{\kappa}\, n \,,
\label{eqn:capacitor}
\end{equation}
Equation~\eqref{eqn:capacitor} is similar to the expression for the voltage on a parallel-plate capacitor with surface charge density $\pm e n$ on the plates and is known as the ``capacitor'' formula~\cite{Ivanov02}. The capacitor formula provides a qualitative explanation for the observed increase of $\Delta E$ with photoexcitation power. However, analytical theory beyond the Hartree approximation~\cite{Yoshioka90, Zhu95, Lozovik97, Ben-Tabou_de-Leon03, Zimmermann07, Schindler08, Laikhtman09, Ivanov10} and Monte-Carlo calculations~\cite{DePalo02, Maezono13} suggest that the capacitor formula significantly overestimates $\Delta E$. The origin of this overestimation is the following. The repulsively interacting IXs avoid each other that lowers their interaction energy per particle as well as their energy shift compared with the uncorrelated state assumed in the Hartree approximation. The correlations can be quantified by the dimensionless correlation parameter $\gamma$ \cite{Remeika09, Remeika15}: $\gamma = \bar\nu_1 \Delta E /\, n\,$, where $\bar\nu_1 = {m}\,/\,{(2\pi \hbar^2)}$ is the bare exciton density of states and $m = m_e + m_h$, $m_e$, and $m_h$, are the effective masses of excitons, electrons, and holes, respectively. The capacitor formula~\eqref{eqn:capacitor} predicts the density-independent correlation parameter $\gamma_\mathrm{c} = \frac{2 d}{a_e}\,\frac{m}{m_e}\,$, which is about $\gamma_\mathrm{c} \approx 7$ for the GaAs CQW structures studied in Refs.~\cite{Remeika09, Remeika15} and the present work. Here $a_e = \hbar^2 \kappa / (m_e e^2)$ is the electron Bohr radius. This gives the IX energy shift
\begin{equation}
\Delta E = \frac{\gamma}{\gamma_\mathrm{c}} \Delta E_\mathrm{c} = \frac{\gamma}{\gamma_\mathrm{c}} \frac{4\pi e^2 d}{\kappa}\, n \, = g n.
\label{eqn:Sigma_naive}
\end{equation}
where we introduce the interaction parameter $g$. For the plate capacitor formula $g_\mathrm{c} = \frac{4\pi e^2 d}{\kappa}$
[Eq.~\eqref{eqn:capacitor}]. Recent experiments \cite{Remeika09, Remeika15} indicated strong correlation in IX system with $\gamma / \gamma_\mathrm{c} \sim 0.1$ at low exciton densities. At the highest electron-hole densities $\gamma / \gamma_\mathrm{c}$ reaches 1. Accurate measurements of $\gamma (n) / \gamma_\mathrm{c}$ yet have to be done. Within the model, the IX energy shift due to interaction is approximated as $\Delta E = g n$ and estimates for different values of $g$ are presented in this work.

(iii) Spinless exciton system is considered within the model. The interaction between different IX spin states as well as accurate values for the splitting of IX spin states yet have to be determined.

(iv) In experimental studies, at low densities and temperatures, IXs in the trap are localized by the in-plane disorder potential (disorder in narrow CQW samples is mainly due to fluctuations of the QW width) \cite{High09PRL}. However, with increasing density, the disorder is screened by exciton-exciton interaction, and, at high densities, the trap behaves as a smooth potential. A smooth trap potential (Fig.~1c) is considered in this work.

To find the exciton energy levels $E_i$ and stationary states $\psi_i$ we solve 2D single-particle Schr{\"o}dinger equation
\begin{equation}
\left( \frac{\hat{p}_x^2+\hat{p}_y^2}{2m}+\hat{V}+ \hat{U}\right) \psi_i = E_i\psi_i. \label{Schrodinger}
\end{equation}
Here, the exciton mass $m = 0.22 m_0$ \cite{Butov01Magn} ($m_0$ is the free electron mass), $\hat{V}$ is the external potential (Fig.~1c), and $U(x,y) = g n(x,y)$ is the interaction potential.

The IX density $n(x,y)$ is calculated using the equilibrium occupation numbers of the single-particle states $N_i$, which are defined by the Bose-Einstein distribution
\begin{gather}
n(x,y) = \sum_{i=0}^\infty N_i \left| \psi_i(x,y) \right|^2,\label{rho} \\
N_i = \frac{1}{\exp\left(\frac{E_i-\mu}{k_\mathrm{B}T}\right)-1}.
\end{gather}
In the last formula, $k_\mathrm{B}$ is the Boltzmann constant, $T$ is the temperature of the exciton gas, $\mu$ is the chemical potential, which depends implicitly on the total number of excitons $N$
\begin{equation}
N = \sum_{i=0}^\infty N_i(\mu,T).\label{N}
\end{equation}

\section{III. Simulation results and discussions}

\begin{figure}
\includegraphics{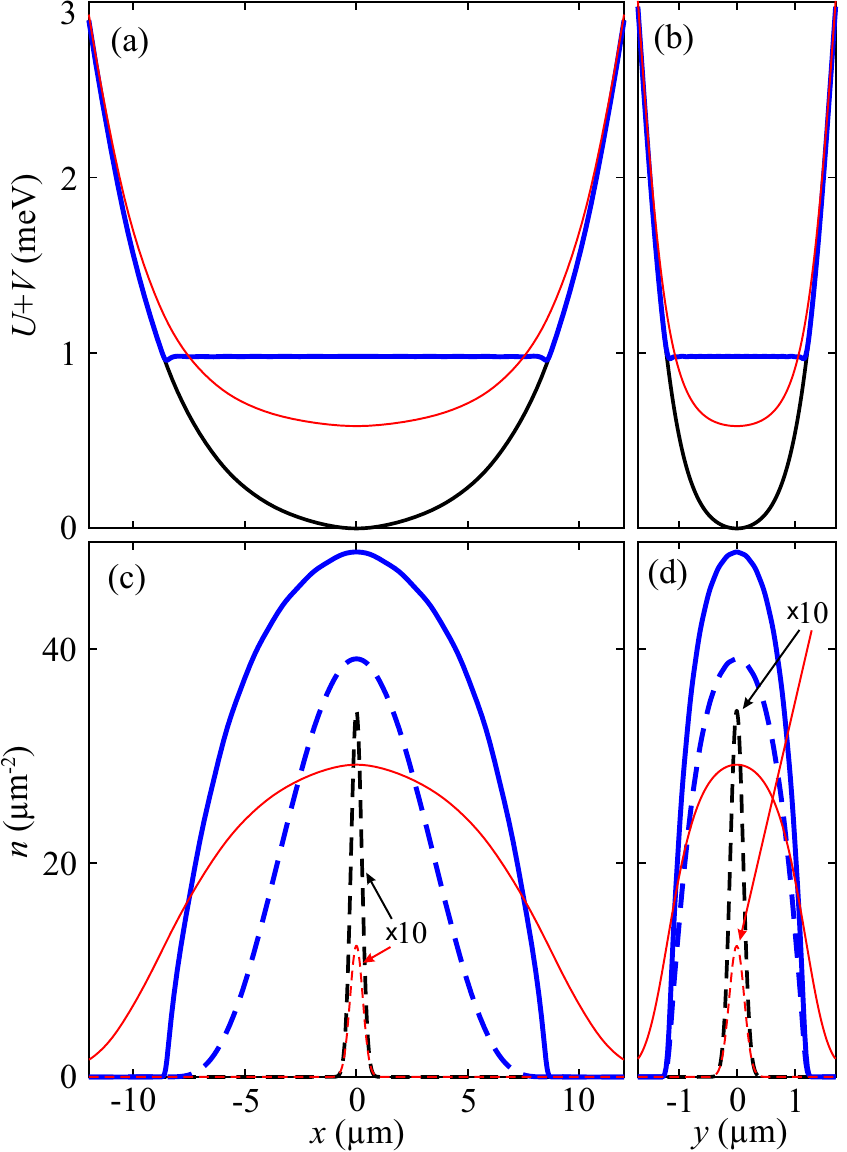} %[width=8.5cm]
\caption{(a,b) The IX energy profile in the bare diamond-shaped trap (black) and screened diamond-shaped trap for $T = 50$~mK (blue thick) and 10~K (red narrow). (c,d)~The corresponding IX density distribution $n$ (solid) and ground state density distributions $n_0$ (dashed). For the screened trap in (a-d), the IX number in the trap $N = 750$, interaction parameter $g = 20$~$\mu\mathrm{eV}\cdot\mu\mathrm{m}^2$.
}
\end{figure}

The bare trap profile (Fig.~1c) is parabolic-like both in $x$ and $y$ directions (Fig.~2a,b, black lines), $V(x,y) \approx \frac{m}{2} (\omega_x^2x^2 + \omega_y^2y^2)$ with $\hbar \omega_x \approx 3$~$\mu$eV and $\hbar \omega_y \approx 14$~$\mu$eV. The effects of screening and condensation of IXs in a trap are demonstrated in Figs.~2--4 for the number of IXs in the trap $N = 750$. At low temperature $T = 50$~mK, IXs effectively screen the trap so the profile of the screened trap is almost flat (Fig.~2a,b, blue lines). At $T = 50$~mK, the exciton density essentially follows the inverted trap profile (Fig.~2c,d, blue lines) that is characteristic for a system with an energy shift due to interaction $\Delta E \propto n$. The width of the ground state density distribution $n_0(x,y)=N_0 | \psi_0(x,y) |^2$ along $x$ and $y$ axes increases by about 13 and 7 times, respectively, due to the trap screening (compare blue and black dashed lines in Fig.~2c,d). Similar features for screening of the trap potential were found for cold atoms in a trap \cite{Dalfovo99, Leggett01}. More than half of IXs are in the ground state $i=0$ at $T = 50$~mK (compare solid and dashed blue lines in Fig.~2c,d). No pronounced bimodal distribution is found for the IX density (Fig.~2c,d, solid blue lines), consistent with a strong repulsive interaction between IXs and, consequently, a relatively wide ground state wave function (Fig.~2c,d, dashed blue lines).

The screening of the trap by IXs is less effective at high temperatures (Fig.~2a,b, red lines). Due to this, the width of the ground state density distribution along $x$ and $y$ axes increases by about 11 and 6 times, respectively, with lowering the temperature (compare blue and red dashed lines in Fig.~2c,d). At the same time, the width of the entire exciton cloud reduces with lowering the temperature due to the IX collection toward the trap center (compare blue and red solid lines in Fig.~2c,d).

\begin{figure}
\includegraphics{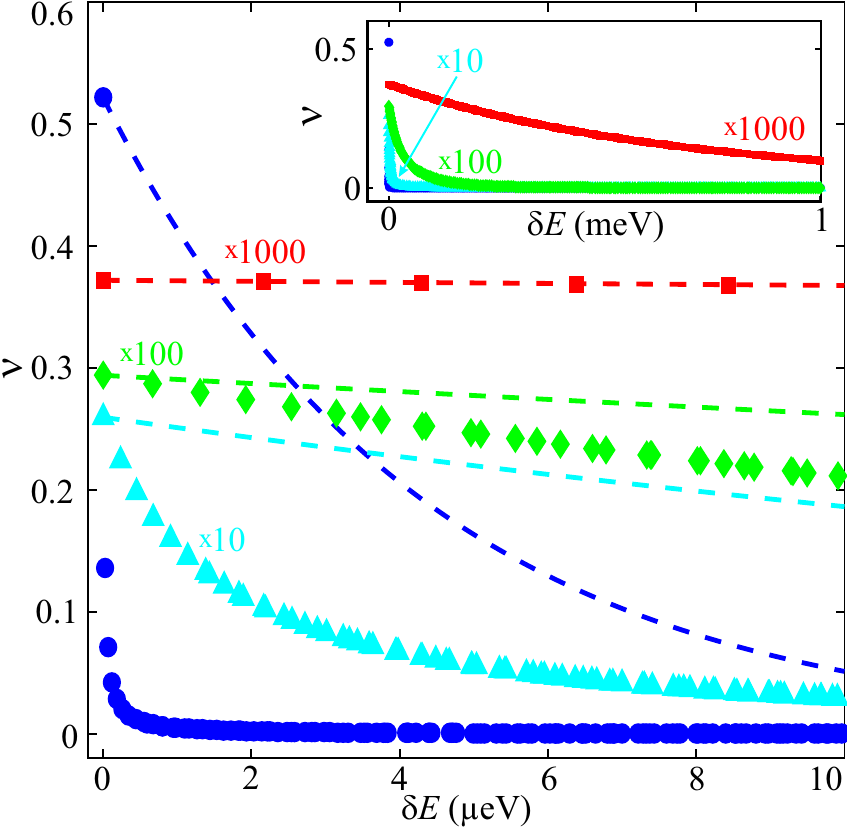}%[width=8.5cm]
\caption{Occupations $\nu_i = N_i/N$ and energies $\delta E_i = E_i - E_0$ of the IX states in diamond-shaped trap for different temperatures $T = 10$~K (red squares, $\times 1000$), 1~K (green diamonds, $\times 100$), 350~mK (cyan triangles, $\times 10$), and 50~mK (blue dots). For the same temperatures, dashed lines show Maxwell-Boltzmann distributions $\propto \exp(\delta E_i/k_\mathrm{B}T)$ brought to $\nu_0$ at $\delta E_i = 0$ for comparison. The inset shows occupations $\nu_i$ for a wider range of energies $\delta E_i$. $N = 750$, $g = 20$~$\mu\mathrm{eV}\cdot\mu\mathrm{m}^2$.
}
\end{figure}

The IX distribution over the trap states is presented for different temperatures in Fig.~3. At high temperature $T = 10$~K, low-energy IX states are nearly equidistant (Fig.~3, red squares) with the splitting between them $\approx 2$~$\mu$eV. With lowering temperature, the trap profile becomes more and more flat and the splitting between IX states reduces to $\approx 0.03$~$\mu$eV at $T = 50$~mK (Fig.~3, blue points, and Fig.~4a, red doted curve).

At high temperature $T = 10$~K, the IX gas is classical with the chemical potential $\mu$ separated from the lowest energy IX state $i=0$ by more than $k_\mathrm{B}T$ (Fig.~4a), the ground state occupation number $N_0 \ll 1$ (Figs.~3 and 4b), and the IX distribution close to the Maxwell-Boltzmann distribution (Fig.~3, compare red squares with red dashed line). For $N = 750$ (Figs.~2--4), the exciton density in the trap reaches $\sim 5 \times 10^9$~cm$^{-2}$ (Fig.~2c,d) and the temperature of quantum degeneracy $T_\mathrm{d} \sim 1$~K. At this temperature, a degenerate Bose gas of IXs forms in the trap with $E_0 - \mu \sim k_\mathrm{B}T$ (Fig.~4a), $N_0 \sim 1$ (Figs.~3 and 4b), and the Bose-Einstein IX distribution (Eq.~5) deviating substantially from the Maxwell-Boltzmann distribution (Fig.~3, compare green diamonds with green dashed line). At low temperature $T = 50$~mK, a Bose-Einstein condensate of IXs forms in the trap with more than half of all IXs in the trap occupying the lowest energy IX state, $N_0 \equiv \nu_0 N \approx 400$ (Figs.~3 and 4b). At this temperature, $E_0 - \mu \ll k_\mathrm{B}T$ (Fig.~4a).

Figure~4b presents the occupation of the ground state $\nu_0 \equiv N_0/N$ and first excited state $\nu_1 \equiv N_1/N$ as a function of temperature. As for any system of bosons of a finite size and finite number of particles \cite{Ketterle86} the condensation transition is smooth so the condensation temperature is somehow uncertain and need to be defined. For certainty, in this work, we define the condensation temperature $T_\mathrm{c}$ as the temperature at which the ground state occupation reaches $1/e$ of all particles in the trap, $\nu_0 \equiv N_0/N = 1/e$ (we note parenthetically that in a 3D infinite system of non-interacting bosons, $1/e$ occupation of the ground state is reached below the condensation temperature $T_\mathrm{c-3D}$, at $T \sim 0.7 T_\mathrm{c-3D}$). This definition results in the condensation temperature $T_c \approx 80$~mK for $N=750$ IXs in the trap for $g = 20$~$\mu\mathrm{eV}\cdot\mu\mathrm{m}^2$ (Fig.~4b). Figure~4b also shows that the occupation of the first excited state $i=1$ reaches maximum around this temperature.

\begin{figure}
\includegraphics{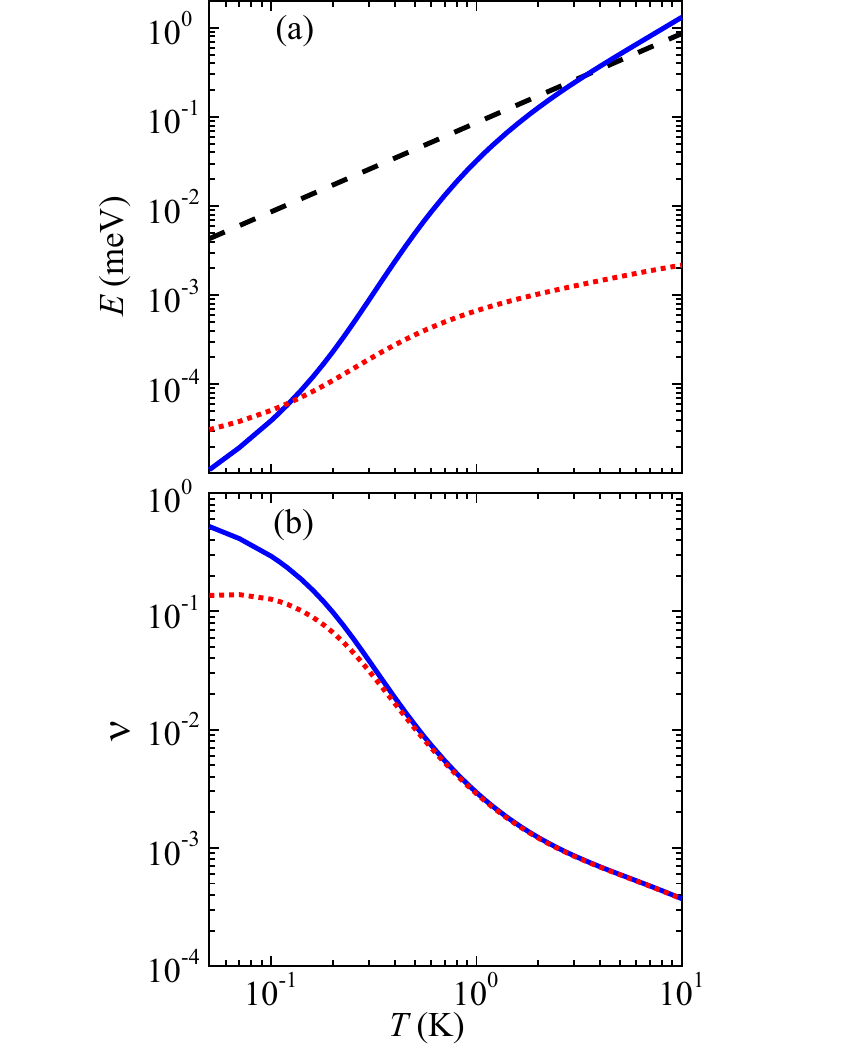} %[width=8.5cm]
\caption{(a) Splitting between the IX ground state $i=0$ and first excited state $i=1$ (red doted) and the IX chemical potential relative to the ground state $E_0 - \mu$ (blue solid) for IXs in diamond-shaped trap as a function of temperature. The thermal energy $k_\mathrm{B}T$ is shown by black dashed line for comparison. (b) The occupation of the ground state $\nu_0 \equiv N_0/N$ (blue solid) and first excited state $\nu_1 \equiv N_1/N$ (red doted) for IXs in diamond-shaped trap as a function of temperature. $N = 750$, $g = 20$~$\mu\mathrm{eV}\cdot\mu\mathrm{m}^2$.
}
\end{figure}

\begin{figure}
\includegraphics{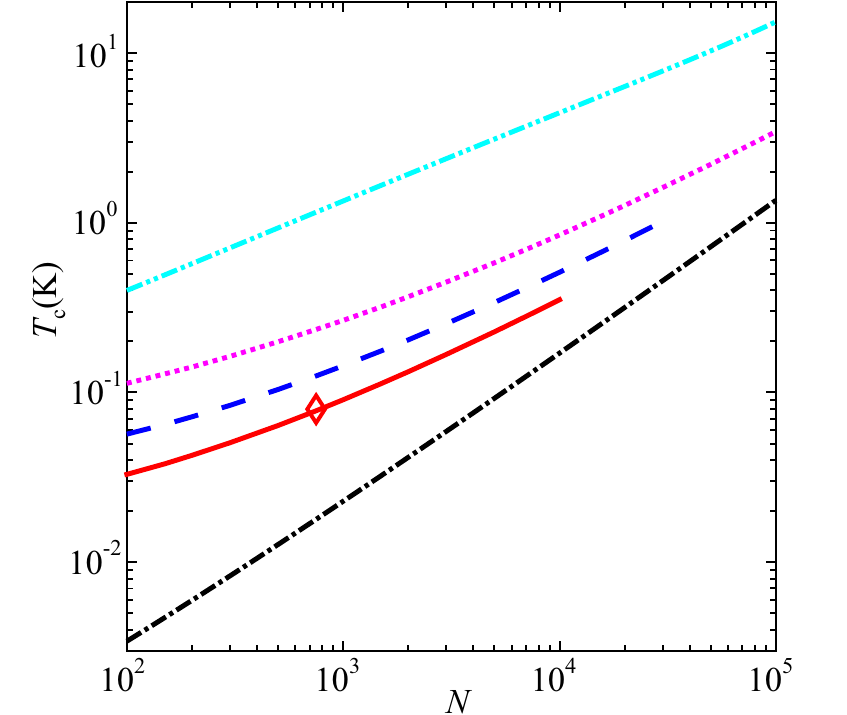}%[width=8.5cm]
\caption{Condensation temperature $T_\mathrm{c}$ vs the number of IXs $N$ calculated for 2D diamond-shaped trap with $g = 20$~$\mu\mathrm{eV}\cdot\mu\mathrm{m}^2$ (red diamond) and for 1D problem with $g = $~20, 7.1, and 2~$\mu\mathrm{eV}\cdot\mu\mathrm{m}^2$ (red solid, blue dashed, and magenta dotted lines, respectively) mapped to the 2D problem as described in the text. $g = 20$~$\mu\mathrm{eV}\cdot\mu\mathrm{m}^2$ corresponds to IX interaction strength given by the "plate capacitor" formula, a lower $g$ corresponds to a weaker IX interaction due to IX correlations as described in the text. The limiting "no screening" case of non-interacting IXs in a parabolic-like trap with $\hbar \omega_x \approx 3$~$\mu$eV and $\hbar \omega_y \approx 14$~$\mu$eV is presented by cyan dash-two-dotted line. The other limit of "complete screening" to flat potential is presented by IXs in $3.5 \times 30$~$\mu\mathrm{m}^2$ rectangular trap with infinite walls (black dash-dotted line).
}
\end{figure}

The density dependence of the IX condensation temperature $T_\mathrm{c}(N)$ is shown in Fig.~5. The limiting cases of a parabolic-like trap with no screening and rectangular trap presenting "complete" screening to a flat box-like potential are presented by the cyan and black lines, respectively. The "parabolic trap" numerical simulation use the parameters of the unscreened trap $V(x,y) = \frac{m}{2} (\omega_x^2x^2 + \omega_y^2y^2)$ with $\hbar \omega_x \approx 3$~$\mu$eV and $\hbar \omega_y \approx 14$~$\mu$eV. The results of these simulations are close (see Fig.~\ref{Tc_vs_N} in Appendix) to the analytical estimates for ideal noninteracting bosons in a parabolic 2D trap $T_\mathrm{c-parabolic} = \frac{\sqrt{6}}{\pi}\hbar \omega_\mathrm{2D} \sqrt{N}$, where $\omega_\mathrm{2D} = \sqrt{\omega_x \omega_y}$ \cite{Ketterle86, Dalfovo99}.

The "rectangular trap" numerical simulations use a rectangular trap potential with the sides equal to the diamond electrode diagonals ($3.5$ and $30$~$\mu\mathrm{m}$). The results of these simulations are close (see Fig.~\ref{Tc_vs_N} in Appendix) to the analytical estimates for ideal noninteracting bosons in a rectangular 2D trap $T_\mathrm{c-rectangular} = T_\mathrm{d}/\ln N = \frac{2 \pi \hbar^2 N}{m A \ln N}$, where $A$ is the area \cite{Ketterle86}. We note that this formula is more accurate for a square trap. For a rectangular trap of a fixed area $A = a_1 a_2$, $T_\mathrm{c-rectangular}$ reduces when the sides of the rectangular $a_1 \ne a_2$ (see Fig.~\ref{Tc_vs_N_shape} in Appendix).

The condensation temperature for the screened diamond trap falls between these limits (Fig.~5, diamond). Since the calculations for a 2D diamond trap require significant computing time, we make the following approximation mapping the 2D problem to 1D problem. The elongated 2D profile of the diamond-shaped trap (Fig.~1c) is replaced by a 1D trap with the same energy profile in the $x$-direction $V=V(x)$. The 2D interaction parameter $g$ is replaced by a 1D interaction parameter $g_\mathrm{1D}$ chosen to produce the same energy shift due to IX interaction in the 1D trap as in the 2D diamond-shaped trap. The corresponding conversion of $g$ to $g_\mathrm{1D}$ is presented in Fig.~\ref{g_1D} in Appendix. The condensation temperature $T_\mathrm{c} = T_\mathrm{c}(N)$ obtained within this approximation is presented in Fig.~5 by red, blue, and magenta lines for different $g$. This 1D approximation produces roughly the same $T_\mathrm{c}$ as the 2D model (compare red line and diamond in Fig.~5).

The highest values of $N$ in Fig.~5 are limited by the IX density at which the trap still provides a confining potential, i.e. is not completely screened. This limiting IX number is determined by the number at which the interaction-induced energy shift is still smaller than the depth of the bare trap potential (Fig.~\ref{E0_vs_N} in Appendix).

Another limit on the IX density is imposed by the Mott transition. The Mott transition is expected when $na^2 \sim 0.1$ (review of the estimates for the Mott transition can be found in Ref.~\cite{Fogler14}). For $a \sim 10$~nm in the structure \cite{Sivalertporn12}, this gives the estimated IX density for the Mott transition $n_M \sim 10^{11}$~cm$^{-2}$. For $N = 750$, the area of the exciton cloud $A \sim 20$~$\mu$m$^2$ (Fig.~2) leading to the estimated number of particles at the Mott transition $N_M \sim A n_M \sim 2 \times 10^4$. The extension of the IX cloud increases with $N$. This is quantified within the 1D approximation by Fig.~6. The cloud extension leads to even higher estimated values for $N_M$, beyond the limits imposed by the complete screening of the trap potential described above.

\begin{figure*}[htb]
\includegraphics{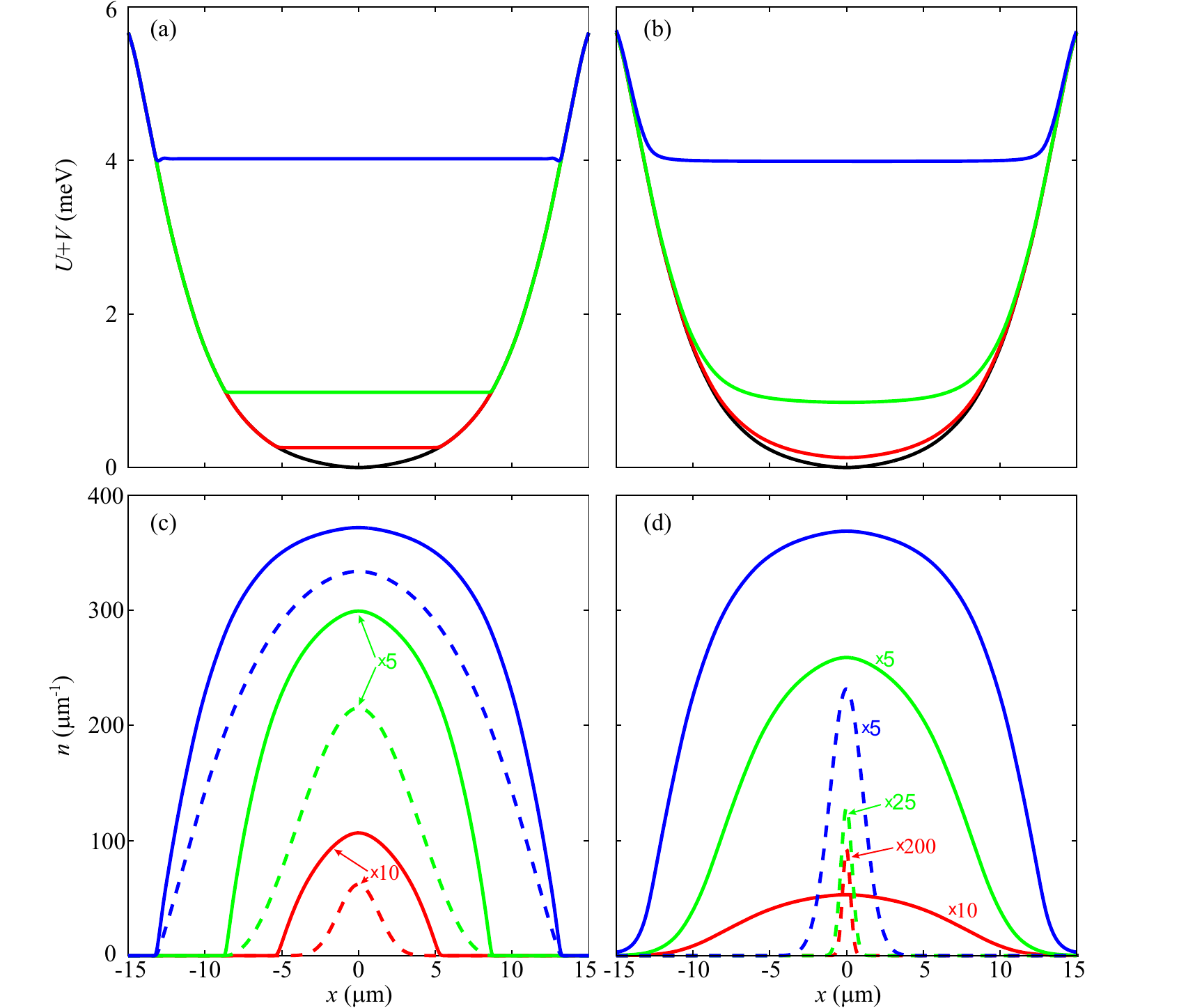}
\caption{(a-b) The IX energy profile in a 1D trap for $T = 50$~mK (a) and 10~K (b). The IX number in the trap $N = 75$ (red), 750 (green) and 7500 (blue). The bare trap energy profile is shown by black line. (c,d) The corresponding IX density distributions $n$ (solid) and ground state density distributions $n_0$ (dashed). For the screened trap in (a-d), the 1D interaction parameters correspond to the 2D interaction parameter $g = 20$~$\mu\mathrm{eV}\cdot\mu\mathrm{m}^2$ as described in the text.
}
\end{figure*}

The red, blue, and magenta lines in Fig.~5 present the simulations for different interaction parameter $g$. Red diamond and red curve in Fig.~5 as well as data in Figs.~2-4 display calculations for $g$, which correspond to IX interaction strength given by the "plate capacitor" formula, $g = 20$~$\mu\mathrm{eV}\cdot\mu\mathrm{m}^2 \sim g_\mathrm{c}$ [Eq.~(1)]. A lower $g$ corresponds to a weaker IX interaction due to IX correlations \cite{Remeika09, Remeika15}. In Fig.~5, the magenta line corresponds to $g \sim 0.1 g_\mathrm{c}$ and the blue line to $g \sim 0.4 g_\mathrm{c}$. Experimental estimates \cite{Remeika09, Remeika15} show that $g$ varies from $\sim 0.1 g_\mathrm{c}$ to $\sim g_\mathrm{c}$ with increasing density. (The accurate measurements of the density dependence for $g=g(n)$ yet need to be done.) This implies that the condensation temperature dependence on $N$ moves from magenta curve at lower $N$ toward blue and red curves at higher $N$.

The simulation results are in qualitative agreement with the experimental data in Ref.~\cite{High12}. In the presence of exciton correlations revealed in earlier measurements of IXs in the studied CQW~\cite{Remeika09, Remeika15}, for the measured IX energy shift 1.3~meV~\cite{High12}, the estimated IX number $N \sim 3 \cdot 10^3 - 10^4$~(blue and magenta lines in Fig.~10) and, in turn, the estimated condensation temperature $T_{\rm c} \sim 0.3 - 1$~K~(Fig.~5). This theoretical estimate is in qualitative agreement with the experimental results: In the experiment~\cite{High12}, with lowering the temperature, the exciton coherence length starts to increase relative to the high-temperature classical value around 2 K and at $T \sim 1$~K the extension of coherence over the entire IX cloud is observed. Note however, that accuracy of the theoretical model should be improved in future works by including to the model the effects of disorder, different spin states, and density- and spin-dependent exciton interaction.

The agreement between the theoretical estimates presented in this work and experimental measurements of IXs condensation in Ref.~\cite{High12} indicates that IX condensation can be adequately described by the theory based on (quasi)equilibrium Bose-Einstein distribution of interacting bosons.

However, it is worth to note the properties of IX systems, which should be taken into account:

(a) The interaction strength between IXs is affected by correlations. The correlations can be estimated theoretically~\cite{Yoshioka90, Zhu95, Lozovik97, Ben-Tabou_de-Leon03, Zimmermann07, Schindler08, Laikhtman09, Ivanov10, DePalo02, Maezono13} and measured experimentally~\cite{Remeika09, Remeika15, Cohen11}. Improving the accuracy in estimates of IX correlations should improve the accuracy of the model of IX condensation.

(b) While at low densities, condensation of the composite bosons - excitons is similar to BEC~\cite{Keldysh68}, at high densities, excitons undergo the Mott transition, above which hydrogen-like excitons dissociate (see~\cite{Fogler14} for review). The theory~\cite{Keldysh65} predicts that a BCS-like exciton condensate may form in electron-hole plasma. The model of IX condensation is applicable for the IX densities below the density of the Mott transition.

(c) The presence of different spin states for excitons may lead to peculiarities of exciton condensation. Some of them are outlined below. IXs in a GaAs CQW may have four spin projections on the $z$ direction: $J_z = - 2 , - 1, + 1, + 2$. The states $J_z = - 1$ and $+ 1$ contribute to left- and right-circularly polarized emission and their coherent superposition to linearly polarized emission, whereas the states $J_z = - 2$ and $+ 2$ are dark~\cite{Maialle93}. (c1) Within the model approximation, the interaction between all states is taken equal. However, the interaction between IXs depends on their $J_z$~\cite{Schindler08}. The accuracy of the model can be improved by taking this into account. Within this approach, the interaction should be affected by the exciton distribution over spin states. This distribution, in turn, depends on the spin state energies. The splitting between the $J_z = \pm 2$ and $\pm 1$ spin states $\Delta$ is determined by the exchange interaction between the electron and hole in the exciton and scales $\Delta \propto \tau_r^{-1}$~\cite{Maialle93, Andreani90, Vinattieri94, Ivchenko05}. For regular direct excitons, DXs, in single GaAs QW, $\Delta < 100$~$\mu$eV~\cite{Ivchenko05}. For IXs in the studied CQW, the radiative lifetime $\tau_r$ is about thousand times larger than for DXs~\cite{Butov99a} and, therefore, the splitting is very small, $\Delta < 0.1$~$\mu$eV. The splitting between the spin states can be also affected by in-plane anisotropy in the structure induced by strain or monolayer fluctuations of interfaces. The measurements of IX spontaneous coherence by shift-interferometry in the region of the rings in exciton emission pattern~\cite{High12N} (without confinement in an electrostatic trap) and the measurements of IX spin polarization by polarization-resolved optical imaging of IX emission~\cite{High13} showed that all four IX spin states form the IX condensate. We note however that other measurements based on the energy shift analysis were discussed in terms of the particle accumulation in the optically dark $J_z = \pm 2$ states~\cite{Shilo13, Cohen16, Alloing14}. (c2) The interaction between IXs also depends on their energies. For instance, the energy dependence of exchange interaction leads to the effective mass renormalization, such renormalization was measured for electron-hole plasma in Refs.~\cite{Butov91, Butov92}. Furthermore, exchange interaction between bosons in the same state vanishes, that can substantially reduce the overall interaction energy at low temperatures where the fraction of IXs occupying the same state is substantial (Fig.~3). The accuracy of the model of IX condensation can be improved by taking these effects into account.

(d) When the distance $d$ between the electron and hole layers is large, interaction between IXs is repulsive and condensation in real space to high-density electron-hole droplets or exciton droplets is energetically unfavorable~\cite{Yoshioka90, Zhu95, Lozovik97}. In turn, in CQW structures with small $d$, condensation in real space may be possible. Condensation in real space was reported in Refs.~\cite{Cohen16, Stern14}. The model of IX condensation is applicable for IXs in CQW (with sufficiently large $d$) where IXs form the ground state (such CQW was studied in Ref.~\cite{High12} and other works).

(e) Optical excitation above the barrier gap, electric leakage currents across the structure, and defects may cause in-plane pattern formations. For instance, electron-rich and hole-rich regions, exciton rings at the interface between these regions~\cite{Alloing14, Butov02, Butov04, Rapaport04, Chen05, Haque06, Yang10, Yang15}, and spatially modulated exciton state~\cite{Alloing14, Butov02, Butov04, Yang15} were observed in gated QW structures. Furthermore, exciton temperature can be higher in the regions of optical excitation~\cite{Hammack06PRL} and in the regions of electric currents through the structure~\cite{Butov04}. The density and temperature inhomogeneities can complicate the condensation. The model of IX condensation is applicable in a system where IX condensation is not affected by IX density or temperature inhomogeneities or free electrons or holes (effects, which may complicate condensation, were avoided in a system of IXs created by optical excitation positioned away from the trap center in the diamond-shaped trap in Ref.~\cite{High12}).

(f) It is worth to mention that there might be other effects, for instance effects related to the device design and performance \cite{device}, which can make the system behavior and data analysis complex.

\section{IV. Summary}

In conclusion, we presented a theoretical model for condensation of indirect excitons in a trap. The model quantifies the effect of screening of the trap potential by IXs on exciton condensation. The theoretical studies were applied to a system of IXs in a GaAs/AlGaAs CQW in a diamond-shaped electrostatic trap where exciton condensation was studied in earlier experiments~\cite{High12}. The qualitative agreement between the theoretical estimates presented in this work and experimental measurements of IXs condensation~\cite{High12} indicates that IX condensation can be adequately described by the theory based on (quasi)equilibrium Bose-Einstein distribution of interacting bosons. The application of the model to various traps and materials, which were studied or can be studied experimentally, as well as improving the accuracy of the simulations by including to the model the effects of disorder, different spin states, and density- and state-dependent exciton interaction form the subject for future work.

\section{Acknowledgements}

This work was supported by NSF Grant No. 1407277 and RFBR Projects No. 14-02-00778 and No. 16-29-03282.

\section{Appendix}

The Appendix presents the comparison of numerical simulations with analytical solutions for condensation temperature $T_\mathrm{c}$ of non-interacting IXs for parabolic and rectangular traps (Fig.~7), the comparison of condensation temperatures $T_\mathrm{c}$ of IXs in rectangular traps with different side ratios (Fig.~8), the relation between the interaction parameters in the 1D and 2D models (Fig.~9), and the IX ground state energy shift vs IX number in the trap (Fig.~10).

\begin{figure}[H]
\centering
\includegraphics[width=7cm]{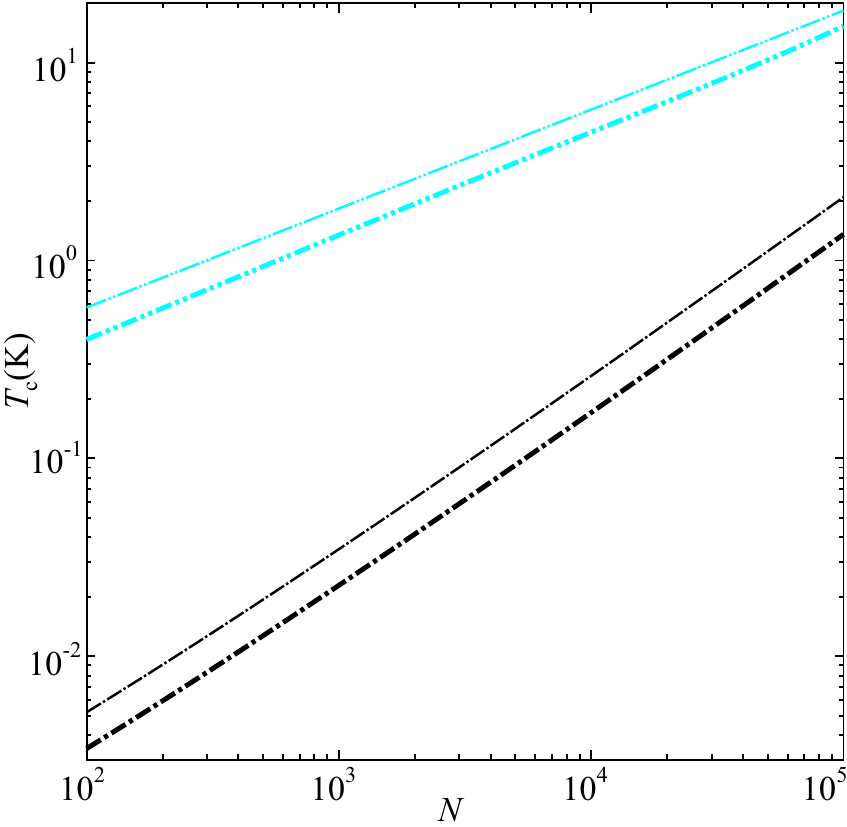}
\caption{Condensation temperature $T_\mathrm{c}$ vs the number of IXs $N$ for analytically solvable problems
of parabolic and rectangular potentials. Numerical simulations for $T_\mathrm{c}$ (cyan dash-two-dotted line) and analytical estimates $T_\mathrm{c-parabolic} = \frac{\sqrt{6}}{\pi}\hbar \sqrt{\omega_x \omega_y N}$ (narrow cyan dash-two-dotted line) for non-interacting IXs in the parabolic potential $V(x,y) = \frac{m}{2} (\omega_x^2x^2 + \omega_y^2y^2)$ with $\hbar \omega_x \approx 3$~$\mu$eV and $\hbar \omega_y \approx 14$~$\mu$eV. Numerical simulations for $T_\mathrm{c}$ (black dash-dotted line) and analytical estimates $T_\mathrm{c-rectangular} = T_\mathrm{d}/\ln N = \frac{2 \pi \hbar^2 N}{m A \ln N}$ (narrow black dash-dotted line) for IXs in the $3.5 \times 30$~$\mu\mathrm{m}^2$ rectangular trap with infinite walls.
}
\label{Tc_vs_N}
\end{figure}

\begin{figure}[H]
\centering
\includegraphics[width=8.5cm]{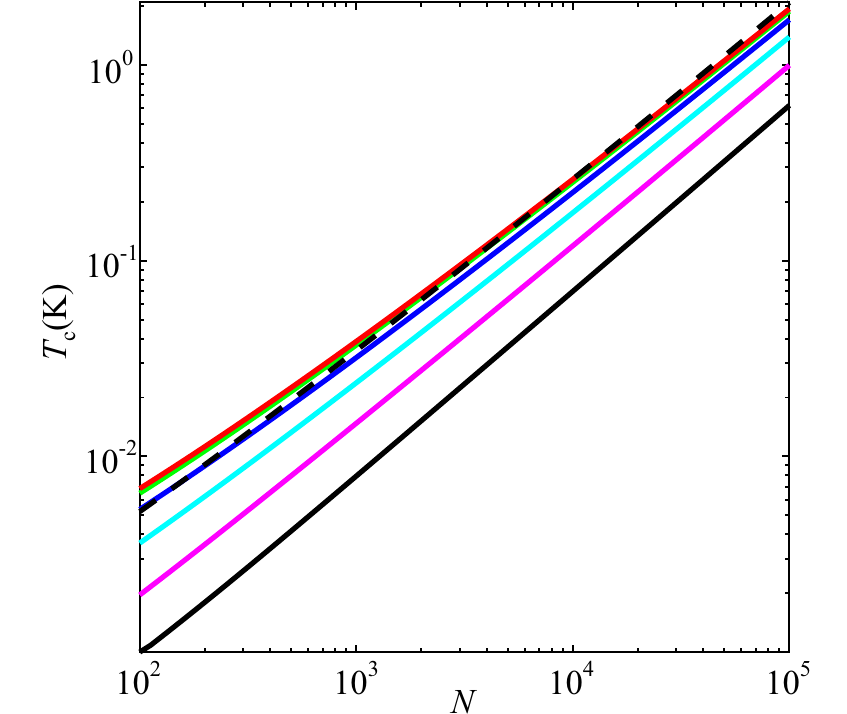}
\caption{Numerical simulations for condensation temperature $T_\mathrm{c}$ vs the number of IXs $N$ for IXs in the rectangular trap with infinite walls for different side ratios $a_1/a_2 = $1, 2, 4, 8, 16, and 32 (red, green, blue, cyan, magenta, and black solid lines, respectively) and the same area $A = a_1 a_2 = 105$~$\mu\mathrm{m}^2$. Black dashed line marks analytical estimate $T_\mathrm{c-rectangular} = T_\mathrm{d}/\ln N = \frac{2 \pi \hbar^2 N}{m A \ln N}$.}
\label{Tc_vs_N_shape}
\end{figure}

\begin{figure}[H]
\centering
\includegraphics[width=8.5cm]{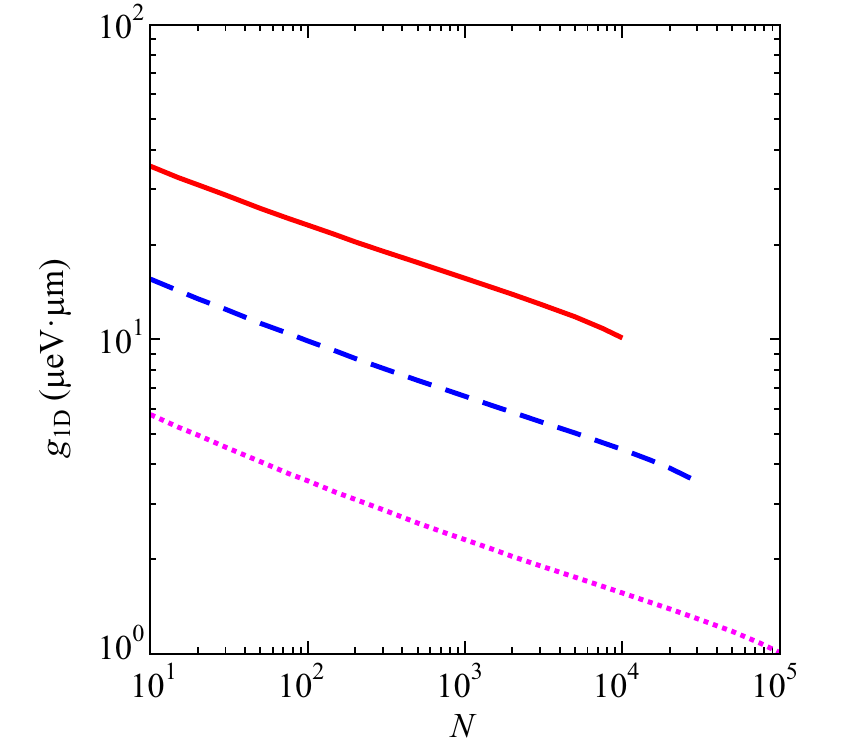}
\caption{One-dimensional interaction parameter $g_\mathrm{1D}$ vs the number of IXs $N$ for three two-dimensional interaction parameters $g = $~2 (magenta dotted), 7.1 (blue dashed), and 20 (red solid)~$\mu\mathrm{eV}\cdot\mu\mathrm{m}^2$. For fixed $N$ and $g$, the 1D interaction parameter $g_\mathrm{1D}$ is such that chemical potentials $\mu$ at $T=0$~K, calculated within the Thomas-Fermi limit (when the kinetic energy term is omitted in Eq.~3), in 1D and 2D problems are the same.
} \label{g_1D}
\end{figure}

\begin{figure}[H]
\centering
\includegraphics[width=8.5cm]{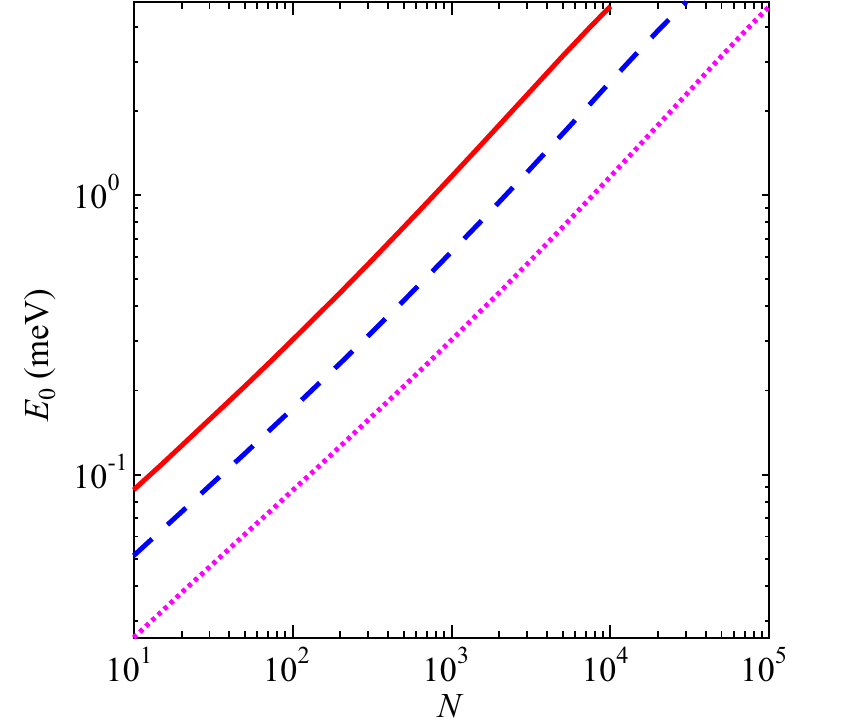}
\caption{The ground state energy $E_0$ vs the number of IXs $N$ for the 1D problem. Three lines (magenta dotted, blue dashed, and red solid) correspond to the 1D interaction parameters $g_\mathrm{1D}(N)$ shown in Fig.~9 by the same colors and, in turn, to 2D interaction parameters $g = 2$ (magenta dotted), 7.1 (blue dashed), and 20 (red solid)~$\mu\mathrm{eV}\cdot\mu\mathrm{m}^2$.
}\label{E0_vs_N}
\end{figure}


\begin{thebibliography}{99}
\bibitem{device}
Electric fields should be properly created in the CQW device. In particular, the top electrodes and bottom homogeneous contact layer should have metallic conductivity, the intrinsic layer between them should be insulating, and the electrodes should have good contact to the semiconductor layers. Otherwise, the applied electric field $F_z(x,y)$ in the device would not correspond to the calculated value (Fig.~1) that would make the model application inaccurate. Furthermore, for imperfect electrodes and bottom contact and insulating layers, their conductivity may depend on position, time, and optical excitation, modifying $F_z(x,y)$ and, as a result, producing IX energy shifts caused by the events in the structure layers surrounding the QWs rather than in IXs.
\end{thebibliography}
\end{document}